\documentclass[12pt,a4paper]{article}
\usepackage{amsfonts,latexsym}
\usepackage{graphicx,color}
\usepackage{dcolumn}
\usepackage{graphicx}
\usepackage{amsmath}
\usepackage{amsfonts}
\usepackage{amssymb}

\renewcommand{\thefootnote}{\fnsymbol{footnote}}

\begin{document}

\vspace{12mm}

\begin{center}
{{{\Large {\bf Scalarization of slowly rotating black holes}}}}\\[10mm]

{Yun Soo Myung$^a$\footnote{e-mail address: ysmyung@inje.ac.kr} and De-Cheng Zou$^{b}$\footnote{e-mail address: dczou@yzu.edu.cn}}\\[8mm]

{${}^a$Institute of Basic Sciences and Department  of Computer Simulation, Inje University Gimhae 50834, Korea\\[0pt] }

{${}^b$Center for Gravitation and Cosmology and College of Physical Science and Technology, Yangzhou University, Yangzhou 225009, China\\[0pt]}
\end{center}
\vspace{2mm}

\begin{abstract}
It is interesting to note that most black holes are born very slowly rotating.
We investigate  scalarization  of slowly rotating  black holes
in the  Einstein-scalar-Chern-Simons (EsCS) theory. In the slow rotation approximation, the CS term  takes a linear form of rotation parameter $a$ which determines the  tachyonic instability.
The tachyonic instability for slowly rotating  black holes represents the onset of spontaneous scalarization.
It is shown  that
the slowly rotating black holes are unstable against a spherically symmetric  scalar-mode
perturbation for  positive coupling $\alpha$, whereas   these black holes are unstable   for  negative coupling without any $a$-bound.

\end{abstract}
\vspace{5mm}

\vspace{1.5cm}

\hspace{11.5cm}
\newpage
\renewcommand{\thefootnote}{\arabic{footnote}}
\setcounter{footnote}{0}


\section{Introduction}
Scalarization of rotating (Kerr) black holes has been  investigated by making use of  the Einstein-scalar-Gauss-Bonnet (EsGB) theory with positive scalar coupling parameter~\cite{Cunha:2019dwb,Collodel:2019kkx}. The tachyonic instability for rotating  black holes is regarded as the onset of  spontaneous scalarization.
In these works, it was noted that  the sufficiently high rotations ($a\ge 0.5$) suppresses scalarization because the GB term is not monotonic and could become negative around the outer horizon.

Recently, the onset of scalarization for  Kerr black holes was found as an $a$-bound of $a/M \ge 0.5$ (sufficiently high rotations) in the EsGB theory with  negative coupling parameter~\cite{Dima:2020yac}.   This $a$-bound  was confirmed   analytically by considering an asymptotic ($l\to\infty$)-scalar mode~\cite{Hod:2020jjy} and numerically by considering the (2+1)-dimensional evolution equation~\cite{Doneva:2020nbb}.
Also, it is shown that any instability could be not  triggered for $a<a_{\rm min}=0.5$ with $M=1$ in the EsGB theory~\cite{Zhang:2020pko}. These imply that the threshold of instability depends on both  the coupling and rotation parameters.
As results,  the spin-induced scalarized  black holes were numerically constructed for $a/M\ge0.5$ in the EsGB theory with negative coupling parameter~\cite{Herdeiro:2020wei,Berti:2020kgk}. Here, we wish to point out that most of spin-induced black hole scalarizations were realized through  scalar coupling to the GB term.

On the other hand, we have found the $a>a(\alpha_{\rm th})$-bound without an $a$-bound when studying  the instability of rotating black holes in the Einstein-scalar-Chern-Simons (EsCS) theory with negative coupling parameter~\cite{Myung:2020etf}. For positive coupling in the EsCS theory, the tachyonic instability for  Kerr black hole were investigated firstly in~\cite{Gao:2018acg}
and its scalarized rotating black holes were constructed~\cite{Doneva:2021dcc}. More recently, the tachyonic instability for Kerr black hole were discussed for a massive scalar in the EsCS theory~\cite{Zhang:2021btn}.

Now, it is  interesting to introduce a slowly rotating black hole because it intrinsically forbids sufficiently high rotations ($a\ge 0.5$).
This black hole could be  found when keeping   all quantities of interest up to first order in $a$ (that is, $a\ll1)$.  At this stage, we would like to mention that
most black holes are born very slowly rotating~\cite{Fuller:2019sxi}. For example, black holes born from single stars rotate
very slowly  with $a=0.01$ and  fairly slow rotating black holes born from single stars  are the cases with $a\le0.1$.  Hence, studying slowly rotating black holes  will reveal  how  $a$ plays the role in achieving spontaneous scalarization for positive and negative coupling parameter $\alpha$.

In this work, we wish to study the onset of scalarization  for  slowly rotating  black holes
in the  EsCS theory with the  coupling parameter $\alpha$.  Its linearized scalar equation includes the CS term depending on $a\cos\theta$.  This may imply no $a$-bound for spontaneous scalarization for negative coupling $\alpha$. If one considers the linearized scalar theory around slowly rotating black holes in the EsGB theory, it involves $48M^2/r^6$ which is independent of $a\cos\theta$, nothing to do with rotation. This is the reason why we consider the EsCS theory for investigating scalarization of slowly rotating black holes. We will employ   the (2+1)-dimensional  hyperboloidal foliation method to show  the tachyonic instability of slowly rotating  black hole by considering the time evolution of a spherically symmetric scalar-mode.

\section{Slowly rotating  black holes }
The EsCS theory takes the form
\begin{eqnarray}
S=\frac{1}{16 \pi}\int d^4 x\sqrt{-g} \Big[
R-\frac{1}{2}(\partial \phi)^2+ \alpha\phi^2~{}^{*}RR\Big] \label{Action}
\end{eqnarray}
with  geometric units of $G=c=1$.
In our model (\ref{Action}), we choose the   quadratic coupling function to  the CS term
\begin{equation}
{}^{*}RR={}^{*}R^{\eta~\mu\nu}_{~\xi}R^\xi_{~\eta\mu\nu}. \label{cs1}
\end{equation}
Here  the dual Ricci tensor is defined as $
{}^{*}R^{\eta~\mu\nu}_{~\xi}=\frac{1}{2}\epsilon^{\mu\nu\rho\sigma}R^{\eta}_{~\xi\rho\sigma}$ with the Levi-Civita tensor $\epsilon^{\mu\nu\rho\sigma}$.
Varying (\ref{Action}) with respect to  $g_{\mu\nu}$ and $\phi$ leads to
 Einstein and scalar equations as
\begin{eqnarray}
&&G_{\mu\nu}=\frac{1}{2}\partial_{\mu}\phi\partial_{\nu}\phi- \frac{1}{4}g_{\mu\nu}(\partial \phi)^2-4\alpha C_{\mu\nu},\label{eqn-1}\\
&&\nabla^2\phi+2\alpha {}^{*}RR\phi=0,\label{eqn-2}
\end{eqnarray}
where $C_{\mu\nu}$ the  Cotton tensor is determined by
\begin{eqnarray}\label{cotton}
C_{\mu\nu}=\nabla_{\rho}(\phi^2)~\epsilon^{\rho\sigma
\gamma}_{~~~~(\mu}\nabla_{\gamma}R_{\nu)\sigma}+\frac{1}{2}\nabla_{\rho}\nabla_{\sigma}
(\phi^2)~\epsilon_{(\nu}^{~~\rho \gamma
\delta}R^{\sigma}_{~~\mu)\gamma \delta}.
\end{eqnarray}

Considering $\phi=0$, Eq. (\ref{eqn-1}) reduces to $R_{\mu\nu}=0$ which implies   the  Kerr spacetime written in Boyer-Lindquist coordinates $\{t,r,\theta,\varphi\}$
\begin{eqnarray}
ds_{\rm Kerr}^2&\equiv&\tilde{g}_{\mu\nu}dx^\mu dx^\nu \nonumber \\
&=& -\frac{\Delta}{\rho^2}(dt-a \sin^2 \theta d\varphi)^2+\frac{\rho^2}{\Delta}dr^2
+\rho^2 d\theta^2 +\frac{\sin^2 \theta}{\rho^2}[a dt -(r^2+a^2)d\varphi]^2   \label{kerr-st}
\end{eqnarray}
with mass ($M$), angular momentum ($J$),  rotation parameter ($a=J/M>0$), $\Delta=r^2-2Mr +a^2$, and $\rho^2=r^2+a^2 \cos^2\theta$.
It is worth noting that Eq. (\ref{kerr-st}) describes a stationary, axisymmetric, and non-static  spacetime.
In this case,  taking into account $\Delta=0$ leads to the outer and inner horizons  as
\begin{equation}
\tilde{r}_\pm=M\Big[1\pm \sqrt{1-\frac{a^2}{M^2}}\Big].
\end{equation}
Before we proceed, we  mention the CS term based on Eq. (\ref{kerr-st}). Its form is given by
\begin{eqnarray}
&&{}^{*}\tilde{R}\tilde{R}=\frac{96rM^2a\cos \theta(3r^4-10r^2a^2\cos^2\theta+3a^4\cos^4\theta)}{\rho^{12}} \nonumber \\
  &&\quad\quad \simeq  \frac{96M^2 a \cos\theta}{r^7} \Big[3-\frac{28a^2\cos^2\theta}{r^2}+\cdots\Big],             \label{CS2}
\end{eqnarray}
where ${}^{*}\tilde{R}\tilde{R}$ is odd with respect to parity transformation:
${}^{*}\tilde{R}\tilde{R}(\pi-\theta)=-{}^{*}\tilde{R}\tilde{R}$. This parity property plays an important role in deriving the threshold curves for negative coupling $\alpha$
when combining with the transformation of $\alpha\to-\alpha$.

From now on, we consider the slowly rotating black hole keeping up to ${\cal O} (a)$-order  in the slow rotation approximation $a\ll 1(J \ll M)$~\cite{Lense:1918zz,Lammerzahl:2018zvb}
\begin{eqnarray}
ds_{\rm SR}^2&=&\bar{g}_{\mu\nu}dx^{\mu}dx^{\nu} \nonumber \\
&=& -\Big(1-\frac{2M}{r}\Big)dt^2+\frac{dr^2}{1-\frac{2M}{r}}
+r^2 (d\theta^2 +\sin^2 \theta d\varphi^2)+\frac{4aM \sin^2\theta}{r} dt d\varphi   \label{srbh}
\end{eqnarray}
which is still a stationary, axisymmetric, and non-static spacetime.
Thus, we  neglect all other terms involving higher order than $a$  in all other quantities of interest: $\bar{R}\simeq0,~\bar{R}_{\mu\nu}\simeq0,~ \bar{R}_{\mu\nu\rho\sigma}\not=0,~\cdots$.
Importantly, the (outer) horizon is   given by the Schwarzschild radius as
\begin{equation}
r_+=2M,
\end{equation}
but the inner horizon is absent.
This implies that the slowly rotating black hole has arisen  from  breaking spherical symmetry to axial symmetry.
Up to ${\cal O} (a)$-order,  the CS term is  given by
\begin{eqnarray}
&&{}^{*}\bar{R}\bar{R}(a)\simeq  \frac{288 M^2 a \cos\theta}{r^7}.       \label{cs2}
\end{eqnarray}
We wish to mention  that Eq. (\ref{cs2}) is a linear term which  approaches zero as $a\to 0$.
Considering Eq. (\ref{cs2}), the time evolution of a spherical mode is determined mainly by $a\cos \theta$, in addition to the coupling parameter $\alpha$.
This means that the rotation $a$ and coupling $\alpha$   play crucial roles in determining the instability bound of slowly rotating black holes, suggesting $\alpha>\alpha_{\rm th}(a)$-bound.

\section{Instability of slowly rotating black holes}
To observe the onset of slowly rotating spontaneous scalarization,
we  introduce the perturbations ($h_{\mu\nu},\delta \phi$)
around the slowly rotating black hole background  as
\begin{eqnarray} \label{m-p}
g_{\mu\nu}=\bar{g}_{\mu\nu}+h_{\mu\nu},~~\phi=\bar{\phi}+\delta\phi \quad {\rm with}~ \bar{\phi}=0.
\end{eqnarray}
The linearized equation to (\ref{eqn-1}) takes a simple form like the general relativity as
\begin{eqnarray}\label{pertg}
\delta R_{\mu\nu}(h)\simeq 0,
\end{eqnarray}
where the linearized Ricci tensor is
\begin{eqnarray}\label{cottonp0}
\delta
R_{\mu\nu}(h)&=&\frac{1}{2}\left(\bar{\nabla}^{\gamma}\bar{\nabla}_{\mu}
h_{\nu\gamma}+\bar{\nabla}^{\gamma}\bar{\nabla}_{\nu}
h_{\mu\gamma}-\bar{\nabla}^2h_{\mu\nu}-\bar{\nabla}_{\mu} \bar{\nabla}_{\nu} h\right).\label{cottonp1}
\end{eqnarray}
The linearized scalar equation around the slowly rotating black hole takes the form
\begin{eqnarray}
 \Big(\bar{\nabla}^2-\mu^2_{\rm CS}\Big)\delta\phi=0,\label{phi-eq2}
\end{eqnarray}
where an effective mass for scalar perturbation is given by
\begin{equation} \label{eff-m}
\mu^2_{\rm CS}=-2\alpha~{}^{*}\bar{R}\bar{R}(a).
\end{equation}
We note that  a tensor-stability analysis for the slowly rotating black hole with Eq. (\ref{pertg}) is the same as in  general relativity,
implying that there are  no unstable tensor modes around the slowly rotating black hole background~\cite{Hafner:2019kov}.
Accordingly, the instability of slowly rotating  black hole will be determined solely by the linearized scalar equation (\ref{phi-eq2}) in the EsCS theory.
For Kerr black hole found from the EsCS theory,  it was shown that the situation with negative $\alpha$ is the same as the case with positive $\alpha$~\cite{Myung:2020etf} because Eq. (\ref{phi-eq2}) is invariant under the transformation of $\alpha\to -\alpha$ and $\theta\to\pi-\theta$~\cite{Zhang:2021btn}.
Initially, we wish to consider two cases of $\alpha>0$ and $\alpha<0$ separately.

\subsection{$\alpha>0$ case}
\begin{figure*}[t!]
   \centering
  \includegraphics{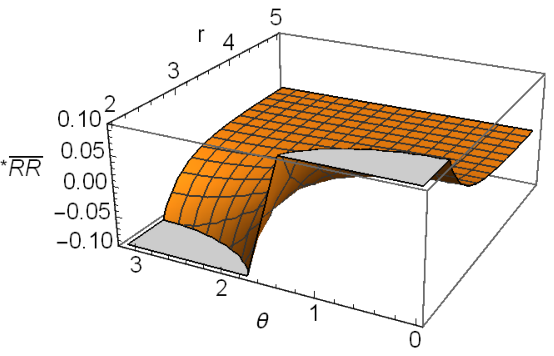}
  \hfill%
  \includegraphics{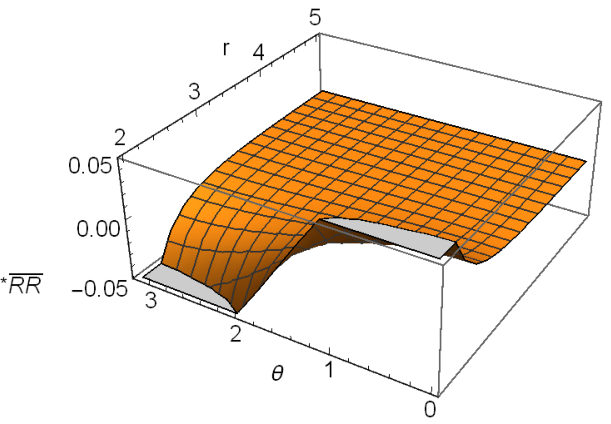}
\caption{3D graphs for CS term ${}^{*}\bar{R}\bar{R}(a)$ with $M=1$ and $a$=0.25 (Left), 0.05 (Right). These include $r\in[r_+=2,5]$ and $\theta\in[0,\pi]$.
Two graphs show division of ($+/-$)-regions near the horizon, implying no distinction in $\mu^2_{\rm CS}$ between positive  and negative $\alpha$ when combining with the transformation of $\alpha\to -\alpha$ and $\theta\to\pi-\theta$.}
\end{figure*}

We find  that  ${}^{*}\bar{R}\bar{R}(a)$  is an odd function with respect to $\cos\theta$.
One finds the division of  ($+$)-region=($-$)-region near the horizon for any $a>0$ (See Fig. 1), implying that it could not distinguish positive coupling from negative coupling~\cite{Gao:2018acg,Myung:2020etf}.
We remind the reader that  the threshold  curve $\alpha=\alpha_{\rm th}(a)$ indicating the boundary between stable and unstable black holes   depends on $a$.
The threshold  curve  will be determined by carrying out numerical simulations. We will take a long time to obtain it.

Let us briefly explain the (2+1)-dimensional hyperboloidal foliation method to solve Eq. (\ref{phi-eq2}) numerically~\cite{Gao:2018acg}.
Firstly, we  introduce the ingoing Kerr-Schild coordinates $\{\tilde{t},r,\theta,\tilde{\varphi}\}$ by considering the coordinate transformations
\begin{equation}
d\tilde{t}=dt+\frac{2 Mr}{\Delta} dr,\quad d\tilde{\varphi}=d\varphi +\frac{a}{\Delta} dr.
\end{equation}
In this case, the linearized scalar equation (\ref{phi-eq2}) could be rewritten  compactly as
\begin{eqnarray}
(\rho^2+2Mr)\partial_{\tilde{t}}^2\delta\phi&=&2M\partial_{\tilde{t}}\delta\phi+4Mr\partial_{\tilde{t}}\partial_r\delta\phi+\partial_r(\Delta\partial_r\delta\phi)
+2a\partial_r \partial_{\tilde{\varphi}}\delta\phi \nonumber\\
&+&\frac{1}{\sin^2\theta}\partial_{\tilde{\varphi}}^2\delta\phi+
          \frac{1}{\sin\theta}\partial_\theta(\sin\theta\partial_\theta\delta\phi)-\mu^2_{\rm CS}\rho^2\delta\phi. \label{phi-eq3} \\
\end{eqnarray}

Considering separation of variables
 \begin{equation}
 \delta \phi(\tilde{t},r,\theta,\tilde{\varphi}) =\frac{1}{r} \sum_{m} u_m(\tilde{t},r,\theta) e^{i m \tilde{\varphi}},
\end{equation}
Eq.(\ref{phi-eq3}) leads to a (2+1)-dimensional Teukolsky equation for a perturbed scalar as
\begin{equation}
A^{\tilde{t}\tilde{t}}\partial_{\tilde{t}}^2u_m+A^{\tilde{t}r}\partial_{\tilde{t}}\partial_ru_m+A^{rr}\partial^2_r u_m+
A^{\theta\theta}\partial_\theta^2u_m+B^{\tilde{t}}\partial_{\tilde{t}}u_m
+B^r\partial_r u_m+B^\theta\partial_\theta u_m+C u_m=0,\label{phi-eq4}
\end{equation}
with coefficients
\begin{eqnarray}
&&A^{\tilde{t}\tilde{t}}=\rho^2+2Mr,~~A^{\tilde{t}r}=-4Mr,~~A^{rr}=-\Delta,~~A^{\theta\theta}=-1,\nonumber\\
&&B^{\tilde{t}}=2M,~~B^r=\frac{2}{r}(a^2-Mr)-2ima,~~B^\theta=-\cot\theta,  \label{coeffs}\\
&&C=\frac{m^2}{\sin^2\theta}-\frac{2(a^2-Mr)}{r^2}+\frac{2ima}{r}+\mu^2_{\rm CS}\rho^2. \nonumber
\end{eqnarray}
As the second step, we wish to solve Eq. (\ref{phi-eq4}) by adopting the hyperboloidal foliation method~\cite{Racz:2011qu}
with   compactified horizon-penetrating hyperboloidal (HH) coordinates $\{\tau, \rho,\theta, \tilde{\varphi}\}$ through
$\tilde{t}=\tau + h(\rho)$ and $r=\rho/\Omega(\rho)$. Here, $h(\rho)=\rho/\Omega-\rho-4M \ln\Omega$ and $\Omega=1-\rho/S$ where $S$  a free parameter determining both the domain and the foliation.  This  implies $\partial_{\tilde{t}}=\partial_{\tau}$ and
$\partial_r=-\frac{dh}{dr} \partial_{\tau}+\frac{d\rho}{dr}\partial_{\rho}$.
The domain  $r\in[r_+,\infty)$ is mapped into a finite region $\rho\in[\rho_+,S)$ with $\rho_+=(a^2S+r_+S^2)/(a^2+2MS +S^2)$.
In this case, Eq. (\ref{phi-eq4}) could be written as
\begin{equation}
\partial^2_{\tau} u_m=\tilde{A}^{\tau\rho}\partial_{\tau} \partial_{\rho} u_m+\tilde{A}^{\rho\rho}\partial^2_\rho u_m+
\tilde{A}^{\theta\theta}\partial_\theta^2u_m+\tilde{B}^{\tau}\partial_{\tau}u_m
+\tilde{B}^\rho\partial_\rho u_m+\tilde{B}^\theta\partial_\theta u_m+\tilde{C} u_m=0, \label{phi-eq5}
\end{equation}
where all coefficients appeared in~\cite{Zhang:2020pko}.
Introducing a momentum $\Pi_m=\partial_{\tau} u_m$,  one finds two coupled first-order equations as
\begin{eqnarray}
\partial_{\tau} u_m&=&\Pi_m, \label{phi-eq6} \\
\partial_{\tau} \Pi_m&=&\tilde{B}^{\tau} \Pi_m+
\tilde{A}^{\tau\rho}\partial_{\rho} \Pi_m +\tilde{A}^{\rho\rho}\partial_{\rho}^2 u_m+ \tilde{A}^{\theta\theta}\partial_{\theta}^2u_m+\tilde{B}^{\rho}\partial_{\rho} u_m+
\tilde{B}^{\theta}\partial_{\theta} u_m+\tilde{C} u_m. \label{phi-eq7}
\end{eqnarray}
The differential equations for  $\rho$ and $\theta$ are solved by using the  finite difference method and the time ($\tau$) evolution is obtained by adopting
the fourth-order Runge-Kutta integrator.
Using the HH coordinates leads to the fact  that the ingoing (outgoing) boundary conditions at the  horizon (infinity)
are satisfied automatically. On the other hand, the boundary conditions at the poles are given as $u_m|_{\theta=0,\pi}=0$ for odd $m=\pm1,\pm3,\cdots$ and
$\partial_{\theta}u_m|_{\theta=0,\pi}=0$ for even $m=0,\pm2,\cdots$.
\begin{figure*}[t!]
   \centering
  \includegraphics{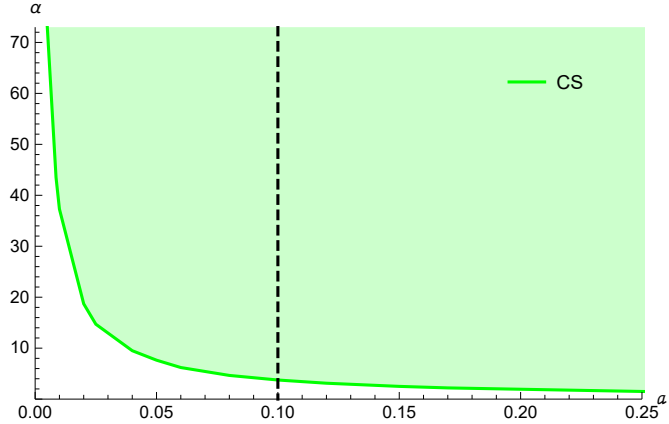}
\caption{CS-threshold (existence) curve $\alpha=\alpha_{\rm th}(a)$ being the boundary between stable  and unstable slowly  rotating black holes is obtained from observing time evolution of a spherically symmetric scalar mode  for positive $\alpha$. The dashed line denotes the upper limit for fairly slow rotating black holes ($0<a\le 0.1$).
}
\end{figure*}

Let us introduce   a Gaussian function [$u_{lm}(\tau=0,\rho,\theta)\sim Y_{lm}(\theta)e^{-\frac{(\rho-\rho_c)^2}{2\sigma^2}}$] localized at $\rho=\rho_c$ outside the horizon as an initial data for a perturbed scalar mode.
Here  $Y_{lm}(\theta)$ denotes the $\theta$-dependent part of spherical harmonics and
$\sigma$ represents the standard deviation. In addition, we impose  the time symmetry  such that $\Pi_{lm}(\tau=0,\rho,\theta)=0$.
Since the slowly rotating spacetime is axisymmetric, the mode coupling may occur.

In this work, we confine ourselves to a spherically symmetric mode of $l=m=0$ for simplicity.
The time  evolution for $\log_{10}|u_{00}(\tau,a,\alpha)|$ provides stable ($\searrow$), threshold ($\longrightarrow$), and unstable ($\nearrow$) case with increasing time ($\tau$).
From Fig. 2, we find  a threshold (existence) curve   $\alpha=\alpha_{\rm th}(a)$  which indicates the boundary between  stable and unstable regions
based on the time evolutions of a scalar mode $\log_{10}|u_{00}(\tau,a,\alpha)|\sim \longrightarrow$.
We observe that the CS-threshold curve decreases  as $a$ increases and it never hits the $\alpha$-axis in the non-rotation limit ($a\to0$).
This curve is  the nearly same as in the curve obtained  with Eq. (\ref{CS2})~\cite{Myung:2020etf}, but one might terminate around $a=0.25$ because of the slow rotation approximation.
The region for fairly slow rotating black holes is given by $0<a\le 0.1$ and the upper limit is represented by a dashed line at $a=0.1$.
The unshaded region [$\alpha<\alpha_{\rm th}(a)$: no growing mode ($\searrow$)] represents stable slowly rotating black holes, while the shaded region [$\alpha\ge \alpha_{\rm th}(a)$: growing mode ($\nearrow$)]  denotes the unstable slowly rotating black holes.
\begin{figure*}[t!]
   \centering
  \includegraphics{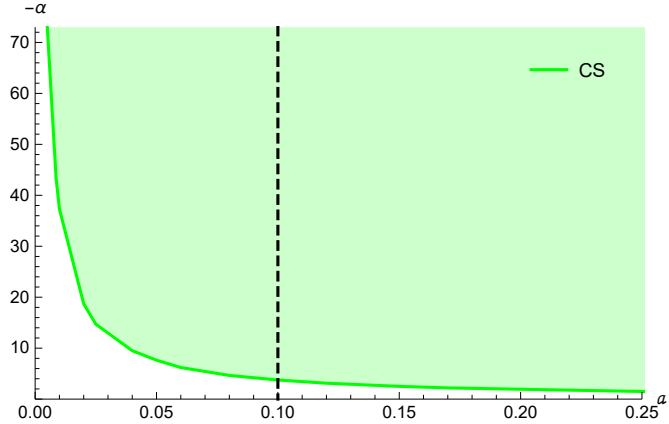}
\caption{ CS-threshold (existence) curve $\alpha=\alpha_{\rm th}(a)$  obtained from observing  the time evolution of $l=m=0$-scalar mode  for negative $\alpha$.
 The dashed line represents the upper limit for fairly slow rotating black holes.}
\end{figure*}

\subsection{$\alpha<0$ case}
In this case, ${}^{*}\bar{R}\bar{R}(a)$ and  $\mu^2_{\rm CS}(a)$ have the same sign.  The whole region near the horizon is always  divided into positive ($+$)-and negative ($-$)-regions because of $\cos \theta$ in the numerator (see Fig. 1), implying no $a$-bound.
The time  evolution for $\log_{10}|u_{00}(\tau,a,-\alpha)|$ implies stable ($\searrow$), threshold ($\longrightarrow$), and unstable ($\nearrow$) with increasing time ($\tau$).
From Fig. 3, we find  a threshold (existence) curve   $\alpha=\alpha_{\rm th}(a)$  which is the boundary between  stable and unstable regions
based on the time evolutions of a scalar mode $\log_{10}|u_{00}(\tau,a,-\alpha)|\sim \longrightarrow$.
Considering the invariance of Eq.(\ref{phi-eq2}) under the transformation  of $\alpha\to -\alpha$ and $\theta\to \pi-\theta$, it is obvious that   the CS-threshold curve (Fig. 3) for  tachyonic instability    is the same  as in Fig. 2~\cite{Myung:2020etf}.  We observe that the CS-threshold curve decreases as $a$ increases and it never hits the $-\alpha$-axis in the non-rotation limit ($a\to0$).
The unshaded region ($\alpha<\alpha_{\rm th}(a)$: no growing mode)  represents stable slowly rotating black holes, while the shaded region ($\alpha\ge \alpha_{\rm th}(a)$: growing mode)  denotes the unstable slowly rotating black holes.

\section{Discussions }
We have investigated spontaneous  scalarization  of slowly rotating  black holes
in the  EsCS theory. The fairly slow rotating black holes imply the cases with $0<a\le 0.1$~\cite{Fuller:2019sxi}.
In the slow rotation approximation with $a\ll1$, the CS term  takes a linear term of $a$-rotation parameter, determining  tachyonic instability.
The tachyonic instability for slowly rotating  black holes represents the onset of spontaneous scalarization.

We have used    the (2+1)-dimensional  hyperboloidal foliation method to show  the tachyonic instability of slowly rotating  black holes when considering a spherically symmetric scalar-mode propagation. The time  evolution for $\log_{10}|u_{00}(\tau,a,\alpha)|$ implies stable ($\searrow$), threshold ($\longrightarrow$), and unstable case ($\nearrow$) with increasing time $\tau$.

It is shown  that
slowly rotating black holes are unstable against a spherically symmetric  scalar-mode of $l=m=0$
for  positive coupling $\alpha$ (Fig. 2). Taking into account the invariance of Eq.(\ref{phi-eq2}) under the transformation  of $\alpha\to -\alpha$ and $\theta\to \pi-\theta$,   the CS-threshold curve (Fig. 3) for  negative coupling  $\alpha$    is the same  as that in Fig. 2~\cite{Myung:2020etf}. This means that   the slowly rotating black holes are  unstable   for  negative coupling $\alpha$ without any $a$-bound.
Finally, it is not difficult to confirm the existence of `no $a$-bound' (like $a>0$) by choosing the $l\to \infty$-scalar mode in the Hod's analytic approach~\cite{Hod:2020jjy,Myung:2020etf}.

 \vspace{1cm}

{\bf Acknowledgments}

 This work was supported by the National Research Foundation of Korea (NRF) grant funded by the Korea government (MOE)
 (No. NRF-2017R1A2B4002057).
 
\end{document}